\documentclass[a4paper,11pt]{article}
\pdfoutput=1 

\usepackage{jcappub} 

\usepackage[T1]{fontenc} 

\title{\boldmath Comparing parametric and non-parametric velocity-dependent one-scale models for domain wall evolution}

\usepackage{graphicx,amssymb}
\usepackage{bbm}
\usepackage[utf8]{inputenc} 

\usepackage{amsmath}

\author[a,b,c,1]{P.P. Avelino\note{Corresponding author.}}

\affiliation[a]{Instituto de Astrof\'{\i}sica e Ci\^encias do Espa{\c c}o, Universidade do Porto, CAUP, Rua das Estrelas, PT4150-762 Porto, Portugal}
\affiliation[c]{Departamento de F\'{\i}sica e Astronomia, Faculdade de Ci\^encias, Universidade do Porto, Rua do Campo Alegre 687, PT4169-007 Porto, Portugal}
\affiliation[c]{School of Physics and Astronomy, University of Birmingham,Birmingham, B15 2TT, United Kingdom}

\emailAdd{pedro.avelino@astro.up.pt}

\abstract{We perform a detailed comparison between a recently proposed parameter-free velocity-dependent one-scale model and the standard parametric model for the cosmological evolution of domain wall networks. We find that the latter overestimates the damping of the wall motion due to the Hubble expansion and neglects the direct impact of wall decay on the evolution of the root-mean-square velocity of the network. We show that these effects are significant but may be absorbed into a redefinition of the momentum parameter. We also discuss the implications of these findings for cosmic strings. We compute the energy loss and momentum parameters of the standard parametric model for cosmological domain wall evolution using our non-parametric velocity-dependent one-scale model in the context of cosmological models having a power law evolution of the scale factor $a$ with the cosmic time $t$ ($a \propto t^\lambda$, $0 < \lambda < 1$), and compare with the results obtained from numerical field theory simulations. We further provide simple linear functions which roughly approximate the dependence of the energy loss and momentum parameters on $\lambda$.}

\begin{document}
\maketitle
\flushbottom

\def\half{{1\over2}}
\def\shalf{\textstyle{{1\over2}}}

\newcommand\lsim{\mathrel{\rlap{\lower4pt\hbox{\hskip1pt$\sim$}}
		\raise1pt\hbox{$<$}}}
\newcommand\gsim{\mathrel{\rlap{\lower4pt\hbox{\hskip1pt$\sim$}}
		\raise1pt\hbox{$>$}}}

\newcommand{\be}{\begin{equation}}
\newcommand{\ee}{\end{equation}}
\newcommand{\bq}{\begin{eqnarray}}
\newcommand{\eq}{\end{eqnarray}}

\section{Introduction}
\label{sec:intr}

The generation of topological defect networks as a result of symmetry breaking phase transitions is a generic prediction of grand unified theories \cite{Kibble:1976sj}. The distinct imprints of cosmic defects on a wide range of cosmological observations constitute a potential window into the fundamental high energy physics of the early universe \cite{Vilenkin:1984ib,2000csot.book.....V,Ade:2013xla}. Domain walls, created as a result of the breaking of a discrete symmetry, are the simplest defect solutions. Unfortunately, the average energy density of standard domain wall networks grows faster than the background density, thus implying that they have to either be extremely light \cite{1974ZhETF..67....3Z} or to have decayed long before the present epoch \cite{Larsson:1996sp,Hindmarsh:1996xv,Avelino:2008qy,Avelino:2008mh,Hiramatsu:2013qaa,Correia:2014kqa,Kitajima:2015nla,Krajewski:2016vbr,Saikawa:2017hiv} in order to be consistent with observations \cite{PinaAvelino:2006ia,Sousa:2015cqa,Lazanu:2015fua}. On the other hand, cosmic strings never tend to dominate the energy density of the universe and, despite the stringent constraints on their tension, they are generally seen as more benign and better motivated than domain walls. The possible production of cosmic superstrings in cosmological scenarios inspired in string theory provides additional theoretical motivation to the observational hunt for cosmic defects \cite{Copeland:2003bj, Dvali:2003zj, Sarangi:2002yt,Sousa:2016ggw}.

The cosmological implications of defect networks depend crucially not only on their microscopic properties but also on their large-scale cosmological dynamics. Although numerical simulations (see, e.g., \cite{Press:1989yh}) are an essential tool towards the understanding of defect network evolution, semi-analytical models are crucial for a deeper understanding of the key dynamical processes responsible for the observational signatures of cosmic defects. A unified framework for the statistical description of the large-scale cosmological evolution of defect networks of relativistic and non-relativistic featureless $p$-branes in $N+1$-dimensional homogeneous and isotropic spacetimes (with $N > p$) was developed in \cite{Avelino:2011ev,Sousa:2011ew,Sousa:2011iu,Avelino:2015kdn}, generalizing previous work on cosmic strings \cite{Martins:1996jp,Martins:2000cs} and domain walls \cite{Avelino:2005kn} (see also \cite{Avelino:2010qf} and \cite{2012PhRvE..86c1119A} for non-cosmological applications). This Velocity-dependent One-Scale (VOS) model provides a characterization of the evolution of the characteristic length $L$ and the root-mean-square velocity $\sigma_v$ of $p$-brane networks, but it relies on the calibration of its two phenomenological parameters, usually referred to as energy loss and momentum parameters, using field theory or Nambu-Goto numerical simulations of cosmic defect network evolution (see also \cite{Martins:2016ois,Rybak2018} for a six parameter extension of the standard parametric domain wall VOS model).

Recently, a VOS model for the cosmological dynamics of standard domain walls free from adjustable dynamical parameters has been proposed \cite{Avelino:2019wqd}. It has been shown to successfully reproduce the results of field theory numerical simulations of domain wall network evolution in cosmological models with a fast expansion rate. The development of this new semi-analytical model has also highlighted a number of problems with the determination of $L$ and $\sigma_v$ from field theory numerical simulations, which will need to be resolved in order for a meaningful comparison to be possible in a relativistic regime. Nevertheless, irrespectively of the exact degree of agreement with numerical simulations, this new domain wall VOS model constitutes an extremely useful tool with which to assess and improve on current semi-analytical models of defect network evolution, an aim which is pursued in the present paper.

The outline of this paper is as follows. In section \ref{sec2} we briefly describe the recently proposed non-parametric VOS model for the cosmological evolution of standard domain wall networks (without junctions). In section \ref{sec3} we compare the VOS equations of motion of  this model with those of the standard parametric VOS model for domain walls, discussing the approximations involved in obtaining the latter from the former. We also briefly discuss the implications of our findings for cosmic strings. In section \ref{sec4} we quantify the impact of these approximations by considering frictionless scaling solutions obtained for a power law evolution of the scale factor with cosmic time. In particular, we compare the values of the energy loss and momentum parameters of the standard parametric VOS model determined using our parameter-free VOS model with the corresponding values obtained using numerical field theory simulations, and provide simple linear functions which approximate the dependence of these two parameters on the expansion rate of the universe. Finally we conclude in section \ref{sec5}.

We shall use fundamental units with $c=1$, where $c$ is the value of the speed of light in vacuum.

\section{Parameter-free VOS model for domain walls}
\label{sec2}

Here, we briefly describe a recently proposed VOS model \cite{Avelino:2019wqd} for the evolution of standard domain wall networks in flat $3+1$-dimensional homogeneous and isotropic Friedmann-Lemaitre-Robertson-Walker (FLRW) cosmologies with line element
\be
d s^2 =a^2[\eta] \left(d \eta^2 - d {\bf x} \cdot d {\bf x}  \right)\,,
\ee
where $a$ is the scale factor, $\eta=\int dt/a$ is the conformal time, $t$ is the physical time and $\bf x$ are comoving spatial coordinates. This model, which is free from adjustable dynamical parameters, describes the cosmological evolution of the network in terms of two thermodynamic variables: the characteristic length $L$ and the root-mean-square velocity $\sigma_v$. It relies on the fact that field theory simulations of standard domain wall network evolution have shown that intersections between domain walls are rare  \cite{Martins:2016ois} (in particular, in comparison to the case of cosmic strings) and that thin domain walls are not expected to produce significant amounts of scalar radiation, except in the final stages of collapse  \cite{Vachaspati:1984yi}. This model makes the simplifying assumption that the universe is permeated by a network of infinitely thin domain walls possessing either spherical or cylindrical symmetry (in the case with cylindrical symmetry the domain walls are taken to be parallel to each other). Also, the gravitational interaction between the walls is assumed to be negligible, thus ensuring that they maintain the initial symmetry throughout the whole evolution and never intersect. Every cosmologically relevant domain wall is assumed to have started at rest at some early conformal time $\eta_i$ with an initial comoving radius $q_i$ much larger than the comoving horizon at that time ($\eta_i$). The probability density function of the initial comoving radii may be written as \cite{Avelino:2019wqd}
\be
{\mathcal P} \equiv {\mathcal P} [q_i] =(1+s) \eta_i^{1+s} q_i^{-2-s} \Theta[q_i-\eta_i] \,,
\ee
where $\Theta$ is the Heaviside step function, and $s=1$ or $s=2$ depending, respectively, on whether the domain walls are assumed to be cylindrical or spherical. The equations of motion describing the frictionless evolution of the domain walls are given by
\be
{\dot q}=-v \,, \qquad v'+(1-v^2)\left(3 {\mathcal H} \eta v - s  \eta  q^{-1}\right)=0\,, 
\label{dwev}
\ee
where a dot denotes a derivative with respect to the conformal time $\eta$, a prime denotes a derivative with respect to $\ln \eta$, ${\mathcal H} \equiv {\dot a}/a > 0$, $q$ is the comoving radius of the wall, $v$ represents its velocity, and $\gamma \equiv (1-v^2)^{-1/2}$. In this model every domain wall is assumed to decay upon reaching $q=0$, and, consequently, we shall only consider the evolution of domain walls in the comoving radius interval $[0,q_i]$. Also, as a consequence of wall decay, at an arbitrary time $\eta \gg \eta_i$ only domain walls with an initial comoving radius larger than a threshold $q_{i*}$ survive, where the value of $q_{i*}$, defined by $q[q_{i*},\eta]=0$, is a function of $\eta$.

The wall energy (spherical case) or energy per unit length (cylindrical case) is equal to $E = 2 \pi s \sigma_{w0} a^s r^s$, where $r=\gamma^{1/s} q$ and $\sigma_{w0}$ is the proper domain wall energy per unit area, or, equivalently,
\be
E=2 \pi s \sigma_{w0} (a \eta)^s \tau^{-s} {\widetilde r}^s\,,
\ee
where ${\widetilde r} \equiv {\widetilde q} \gamma^{1/s}$, ${\widetilde q} \equiv q/q_i$, and $\tau \equiv \eta/q_i$. The evolution of $E$ can be readily computed from the equations of motion and it is given by
\be
E'={\mathcal H} \eta E \left(s-3 v^2\right)\,.
\label{Eev}
\ee

Hence, in our model the average energy density of the domain wall network at the conformal time $\eta$ is equal to
\bq
\rho_{w}&=& \int _{q_{i*}}^\infty n E {\mathcal P}  dq_i  = a^{-1-s} \int _{q_{i*}}^\infty n_i E {\mathcal P}  dq_i \nonumber \\
&=& \sigma_{w0} \beta (a  \eta)^{-1} \int_0^{\tau_*} {\widetilde r}^s d \tau\,, \label{rhow}
\eq
where $\beta=2 \pi s (1+s)  n_i \eta_i^{1+s}$, $n_i$ is the initial domain wall number density defined as the number of walls per unit volume (spherical case) or per unit area (cylindrical case), $n= n_i a^{-1-s}$ would be the domain wall number density at the time $\eta$ in the absence of decay, and $\tau_*=\eta/q_{i*}$. 

Let us also define mean square velocity of the domain walls as
\bq
\sigma_v^2 &=& \langle v^2 \rangle = \frac{\int v^2 \rho \, dV}{\int \rho \, dV}=  \frac{\int v^2  \gamma dS}{\int  \gamma dS}= \frac{\int _{q_{i*}}^\infty v^2 E {\mathcal P}  dq_i }{\int _{q_{i*}}^\infty E {\mathcal P}  dq_i } \nonumber\\
&=&\frac{\int_0^{\tau_*} v^2 {\widetilde r}^s d \tau}{\int_0^{\tau_*} {\widetilde r}^s  d \tau}\,, \label{sigma2}
\eq
where $\rho$ is the (microscopic) domain wall energy density at each point, $V$ is the physical volume, and $S$ is the domain wall area. The third equality in eq. (\ref{sigma2}) is obtained by writing $dV=dSdl$ and performing the integration $\int \rho \, dl = \sigma_w=\sigma_{w0} \gamma$ in the direction perpendicular to the domain wall (note that $\rho=\rho_0 \gamma^2$ and $\delta=\delta_0/\gamma$, where the subscript `$0$' represents the proper rest value and $\delta$ denotes the domain wall thickness, and that $v$ does not vary along the direction perpendicular to the wall).

For a power law expansion, with $a \propto t^\lambda \propto \eta^{\lambda/(1-\lambda)}$ and ${\mathcal H} \eta=\lambda/(1-\lambda)$, the equations of motion are invariant with respect to the transformation $q \to \alpha q$, $\eta \to \alpha \eta$, where $\alpha > 0$ is a constant. In this case $v[q_i,\eta]=v[1,\eta/q_i] = v [\tau]$, $ {\widetilde q}=q[q_i,\eta]/q_i = q[1,\eta/q_i]/q_i =  {\widetilde q}[\tau]$, and $\tau_* = \eta/q_{i*} = {\rm const}$. However, in general ${\widetilde q}[q_i,\eta]={\widetilde q}[\tau,\eta]$, $v[q_i,\eta]=v[\tau,\eta]$ and $\tau_* = \tau_*[\eta]$. 

Differentiating eq. (\ref{rhow}) with respect to $\ln \eta$ and using eq. (\ref{Eev}) one obtains
\bq
\rho_w' &=& - (1+s) {\mathcal H} \eta \rho_w + a^{-1-s}  \int _{q_{i*}}^\infty n_i E' {\mathcal P}  dq_i \nonumber \\ 
&-& a^{-1-s} n_i E_* {\mathcal P} _* q_{i*}' = - (1+3 \sigma_v^2) {\mathcal H} \eta \rho_w \nonumber \\
&-&  \sigma_{w0} \beta (a \eta) ^{-1} {\widetilde r}_*^s  \tau_* \left(1-\tau_* \tau_*' \right) \label{rhow1}\,,
\eq
where $E_*=E[q_{i*},\eta]$, ${\mathcal P}_*={\mathcal P}_*[q_{i*}]$, and ${\widetilde r}_*={\widetilde r}[q_{i*},\eta]$. The last term in eq.  (\ref{rhow1}),
\be
\rho_{w[{\rm decay}]}' = -  \sigma_{w0} \beta (a \eta) ^{-1} {\widetilde r}^s_*  \tau_* \left(1-\tau_* \tau_*' \right) \label{rhowc}\,,
\ee
is associated to the energy losses by the network due to domain wall decay.

On the other hand, differentiating eq. (\ref{sigma2}) with respect to $\ln \eta$ one obtains
\bq
(\sigma_v^2)' &=& - 6 {\mathcal H} \eta \sigma_v^2 \left(1-\frac{\sigma_v^2}{2} - \frac{\langle v^4 \rangle}{2\sigma_v^2}\right)+2 s \eta \left \langle  \frac{v (1-v^2)}{q} \right \rangle \nonumber \\
&-& \left(1-\sigma_v^2\right)   \frac{{\widetilde r}^s_* \tau_* \left(1-\tau_* \tau_*' \right)}{\int_0^{\tau_*} {\widetilde r}^s d \tau} \,, \label{sigma2ev}
\eq
with the averages denoted by $\langle ... \rangle$ being defined as in eq. (\ref{sigma2}), where the specific case of $\langle v^2 \rangle$ was considered. We have also taken into account that $v \to 1$ as $q \to 0$, or, equivalently, that $v_* \equiv v[q_{i*},\eta]=1$.

\section{Relation to the standard parametric VOS model}
\label{sec3}

A parametric VOS model providing a description of the averaged large-scale dynamics of thin and featureless cosmic defect networks in homogeneous and isotropic backgrounds was developed in \cite{Avelino:2011ev,Sousa:2011ew,Sousa:2011iu,Avelino:2015kdn} (see also \cite{Martins:1996jp,Martins:2000cs} and \cite{Avelino:2005kn} for previous work on cosmic strings and domain walls, respectively). This model provides a characterization of cosmic defect network evolution in terms of two dynamical variables, the characteristic length $L$ and root-mean-squared velocity $\sigma_v$. 
	
In the case of domain walls, the characteristic length of the network is defined as 
\be
L \equiv \frac{\sigma_{w0}}{\rho_w}\,. \label{Ldef}
\ee
The corresponding VOS equations of motion are given by \cite{Avelino:2011ev,Sousa:2011ew,Sousa:2011iu}
\bq
\zeta ' &=& -  \zeta \left(1-3 {\mathcal H} \eta   \sigma_v^2\right) + c_w  \sigma_v  \label{vos1a}\,,\\
\sigma_v ' &=& - (1-\sigma_v^2) \left(  3  {\mathcal H} \eta \sigma_v - k_w \zeta^{-1} \right) \label{vos2a}
\eq
where $k_w={\bar \kappa}_w L/a$,
\be
{\bar \kappa}_w=\frac{\langle v(1-v^2)\kappa_w\rangle }{\sigma_v(1-\sigma_v^2)}\,,
\ee
$\kappa_w$ is equal to the sum of the two comoving principal curvatures at each point on the wall ($\kappa_w=s/q$ with $s=1$ or $s=2$ for cylindrical or spherical domain walls, respectively), and $c_w$ is a phenomenological energy loss parameter defined by 
\be
\rho_{w[{\rm decay}]}' =-c_w \sigma_v\zeta^{-1} \rho_w\,.
\ee
In the derivation of eqs. (\ref{vos1a}) and  (\ref{vos2a}) from the microscopic equations of motion it has been assumed that $\langle v^4 \rangle =\sigma_v^4$. Also, the standard parametric VOS model assumes that $k_w$ can be approximated as a function of $\sigma_v$ alone [$k_w=k_w(\sigma_v)$].

In the non-parametric  VOS model eq. (\ref{Ldef}) in combination with eq. (\ref{rhow}) implies that
\be
\zeta \equiv \frac{L}{a\eta}=\left(\beta \int_0^{\tau_*} {\widetilde r}^s  d \tau\right)^{-1} \label{zetadef}\,.
\ee
Using eqs. (\ref{rhow1}), (\ref{sigma2ev}), (\ref{Ldef}), and (\ref{zetadef}) it is possible to cast the equations of motion of our parameter-free VOS model in a similar form to that of the parametric VOS model, that is
\bq
\zeta ' &=& -  \zeta \left(1-3 {\mathcal H} \eta   \sigma_v^2\right) + c_w  \sigma_v  \label{vos1}\,,\\
\sigma_v ' &=& - (1-\sigma_v^2) \left(  3 (f+1) {\mathcal H} \eta \sigma_v - k_w \zeta^{-1} + \epsilon\right) \label{vos2}
\eq
with
\bq
k_w &=&  \frac{s \zeta \tau \langle v (1-v^2) {\widetilde q}^{-1}\rangle}{\sigma_v (1-\sigma_v^2) } \,, \label{kw}\\
c_w &=& 2 \epsilon \zeta \,, \label{cw}\\
\epsilon&=&\frac{\beta}{2} \frac{\zeta}{\sigma_v} {\widetilde r}^s_*\tau_* \left(1-\tau_* \tau_*' \right)\,,\\
\chi &=&  \frac12 + \frac12 \frac{\langle v^4 \rangle}{\sigma_v^4}\, \qquad f=\frac{\sigma_v^2(1-\chi)}{1- \sigma_v^2} \,.
\eq
Equation (\ref{vos1}) may also be written as 
\bq
(\ln \zeta) ' &=& -  \left(1-3 {\mathcal H} \eta   \sigma_v^2 -2 \epsilon \sigma_v\right) \label{vos3}\,.
\eq
The right-hand sides of eqs. (\ref{vos2}) and (\ref{vos3}) are independent of $\beta$ (note that eq. (\ref{zetadef}) implies that $\beta \zeta$ is independent of $\beta$), thus implying that for a given cosmological model the evolution of $\ln \zeta$ and $\sigma_v$ is completely determined by our model both for $s=1$ and $s=2$, without the need for adjustable phenomenological dynamical parameters.

It is useful to compare eqs. (\ref{vos1}) and (\ref{vos2}) with those of the standard parametric VOS model for domain walls in which $f$ is assumed to be equal to zero (the same approximation is also made in the standard parametric VOS model for cosmic strings). Taking into account that $\langle (v^2-\sigma_v^2)^2 \rangle = \langle v^4 \rangle - \sigma_v^4 \ge 0$, it is simple to show that
\be
f=-\frac{\langle (v^2-\sigma_v^2)^2 \rangle}{2 (1-\sigma_v^2)\sigma_v^2} \le 0\,
\ee
Hence, $f$ is equal to unity only if there is no dispersion of the root-mean-square velocity, which in general may only happen in the non-relativistic ($\sigma_v \to 0$) or in the ultra-relativistic limit ($\sigma_v \to 1$). In the following section we shall determine the values  of $f$ predicted by our parameter-free domain wall VOS model for different values of the expansion rate.

\begin{figure}[!htb]
	\centering
	\includegraphics[width=12.0cm]{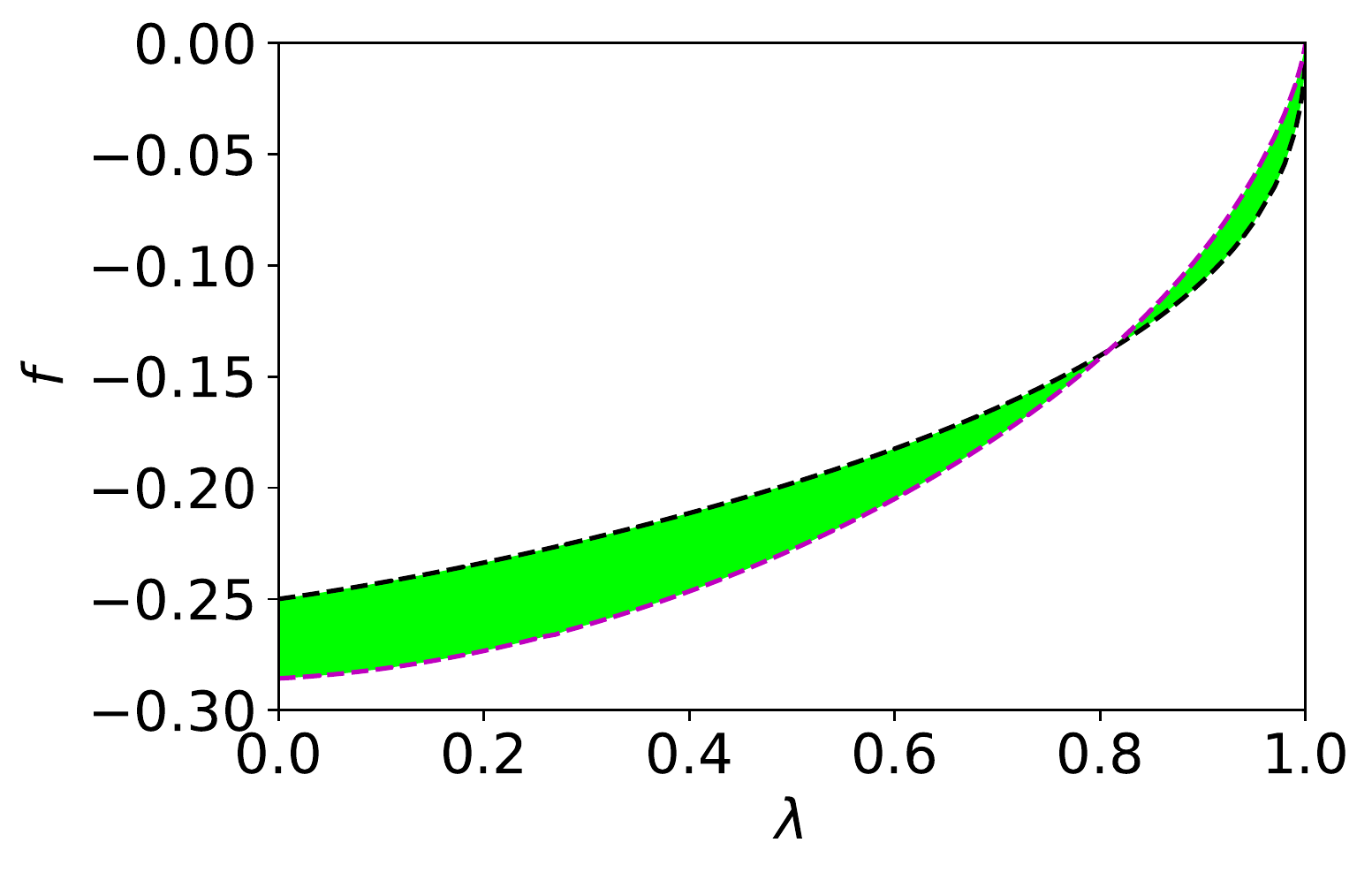}
	\caption{The value of $f$ predicted by our parameter-free VOS model as a function of $\lambda$ in the context of frictionless scaling solutions considering cylindrical or spherical domain walls (upper black and lower magenta dashed lines, respectively) --- the green region between the two lines provides an estimate of the model uncertainty. Notice that the $f=1$ approximation of the parametric VOS model for domain walls overestimates the strength of the Hubble damping due to the expansion of the universe.}
	\label{fig1}
\end{figure}

On the other-hand, the direct contribution of $\epsilon$ in eq. (\ref{vos2}) is not explicitly considered in the standard parametric VOS model. In practice such contribution, is equivalent to the following redefinition of the momentum parameter: $k_w \to k_w - \epsilon \zeta$. In the standard parametric VOS model it is implicitly assumed that the elements removed from the network --- in the case of domain walls as a consequence of collapse and decay --- have a root-mean-square velocity equal to that of the network (thus implying that $\epsilon=0$). However that is not in general the case. Whenever there is a difference between the root-mean-square velocity of the network and that of the elements being removed from it, there will be a direct contribution to the change in $\sigma_v$ given by
\be
d (\sigma_v^2) =-  \frac{d \rho_{\mathcal N}}{\rho_{\mathcal N}} \left(\sigma_v^2 - \sigma_{v-}^2\right)\,,
\ee
where the subscripts $\mathcal N$ and $-$ represent, respectively, the network and the elements being removed from it. In the case of domain walls the velocity is extremely close to unity prior to decay and, therefore, 
\be
\sigma_{v[\rm decay]}' = \frac{\rho_{w[{\rm decay}]}'}{2\rho_w} \frac{1-\sigma_v^2}{\sigma_v}=-(1-\sigma_v^2) \epsilon\,,
\ee
where we have taken $\sigma_{v-}=1$.

To our knowledge, this effect has not been considered in previous work, not only in the context of domain walls but also of cosmic strings. For cosmic strings one would expect a contribution to the change of $\sigma_v$
\be
\sigma_{v[\rm loops]}' = \frac{\rho_{str[{\rm loops}]}'}{\rho_{str}} \frac{\sigma_{v-}^2- \sigma_v^2}{2\sigma_v}= \frac{ \sigma_v^2-\sigma_{v-}^2}{2}c_{str} \zeta^{-1} \,,
\ee
coming from loop production. Here, $\rho_{str}$ is the average string energy density, $c_{str}$ is the energy loss parameter of the cosmic string VOS model, and $\sigma_{v-}$ is the root-mean-square velocity of the loops produced by the long string network in the cosmological frame.

\begin{figure}[!htb]
	\centering
	\includegraphics[width=12.0cm]{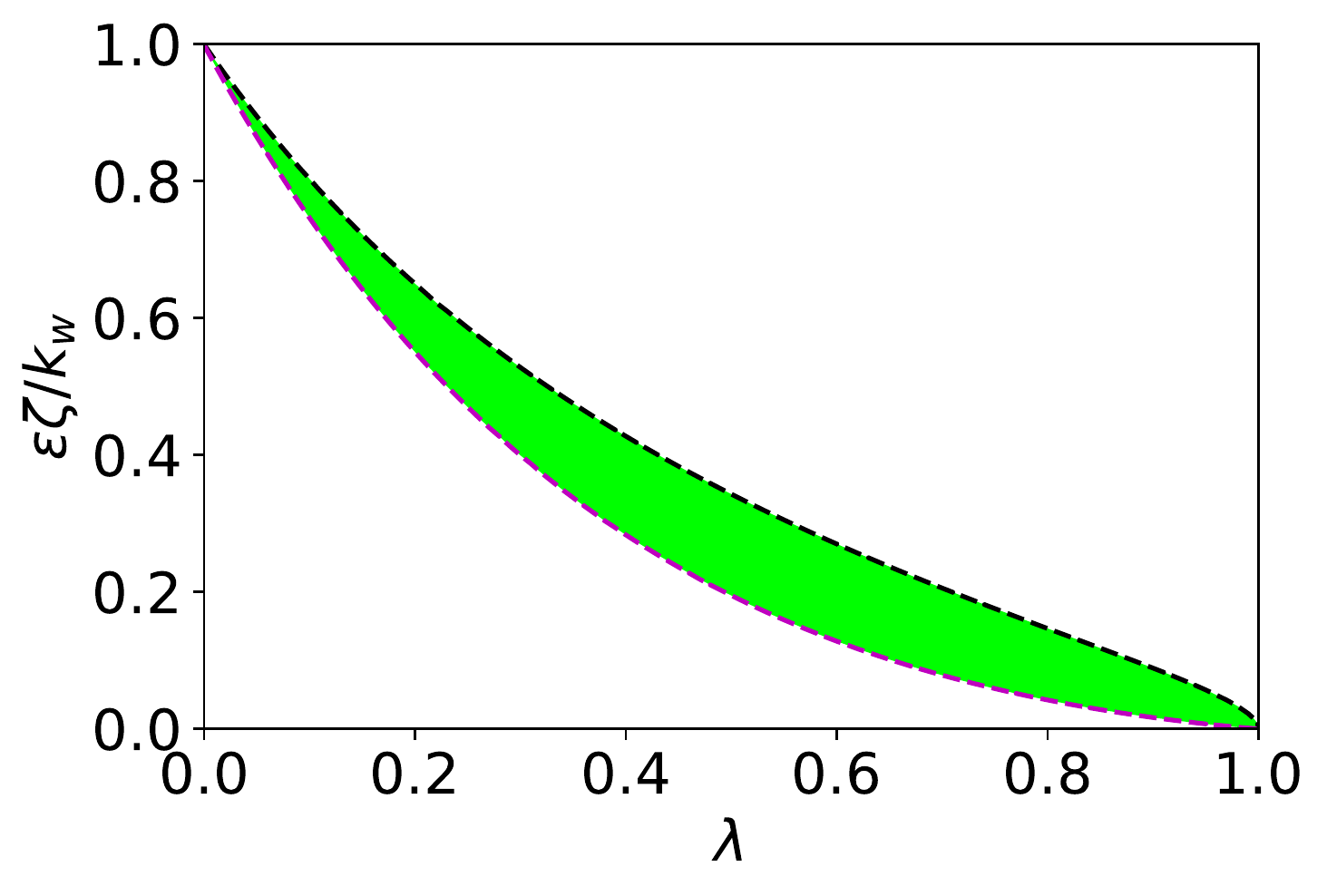}
	\caption{The value of $\epsilon \zeta/ k_w$ predicted by our parameter-free VOS model as a function of $\lambda$ in the context of frictionless scaling solutions considering cylindrical or spherical domain walls (upper black and lower magenta dashed lines, respectively) --- the green region between the two lines provides an estimate of the model uncertainty. Notice that, except for values of $\lambda$ close to unity, the last two terms in eq. (\ref{vos2}) are always of the same order of magnitude.}
	\label{fig2}
\end{figure}

\section{VOS model for domain walls: frictionless scaling solutions}
\label{sec4}

Here, we shall consider frictionless scaling solutions in the context of cosmological models having a power law evolution of the scale factor with the physical time ($a \propto t^\lambda$, $0<\lambda <1$), so that $\zeta$ and $\sigma_v$ are constants. Figure \ref{fig1} displays the values of $f$ predicted by our parameter-free VOS model, as a function of $\lambda$, considering cylindrical or spherical domain walls (upper black and lower magenta dashed lines, respectively). The green region between the two lines provides an estimate of the model uncertainty associated to the geometry of the domain walls. Figure \ref{fig1} shows that, except for values of $\lambda$ close to unity, $f$ deviates significantly from zero. It implies that the $f=0$ assumption made in the standard parametric VOS model for domain walls leads to an overestimation of the strength of the Hubble damping by $20 \pm 10 \%$ for values of $\lambda$ in the interval $[0,0.9]$. A similar effect is also expected in the standard parametric cosmic string VOS model, albeit with a slightly smaller amplitude.

The value of $\epsilon \zeta/ k_w$ predicted by our parameter-free VOS model as a function of $\lambda$ considering cylindrical or spherical domain walls is shown in Fig.  \ref{fig2} (upper black and lower magenta dashed lines, respectively, with the green region between the two lines again representing an estimate of the model uncertainty). It  parameterizes the relative importance of the last two terms in eq. (\ref{vos2}). Figure \ref{fig2} shows that for values of $\lambda$ not too close to unity these two terms are of the same order of magnitude, thus implying that the direct impact of domain wall decay in eq. (\ref{vos2}), associated to a non-zero $\epsilon$, cannot in general be neglected.

\begin{figure}[!htb]
	\centering
	\includegraphics[width=12.0cm]{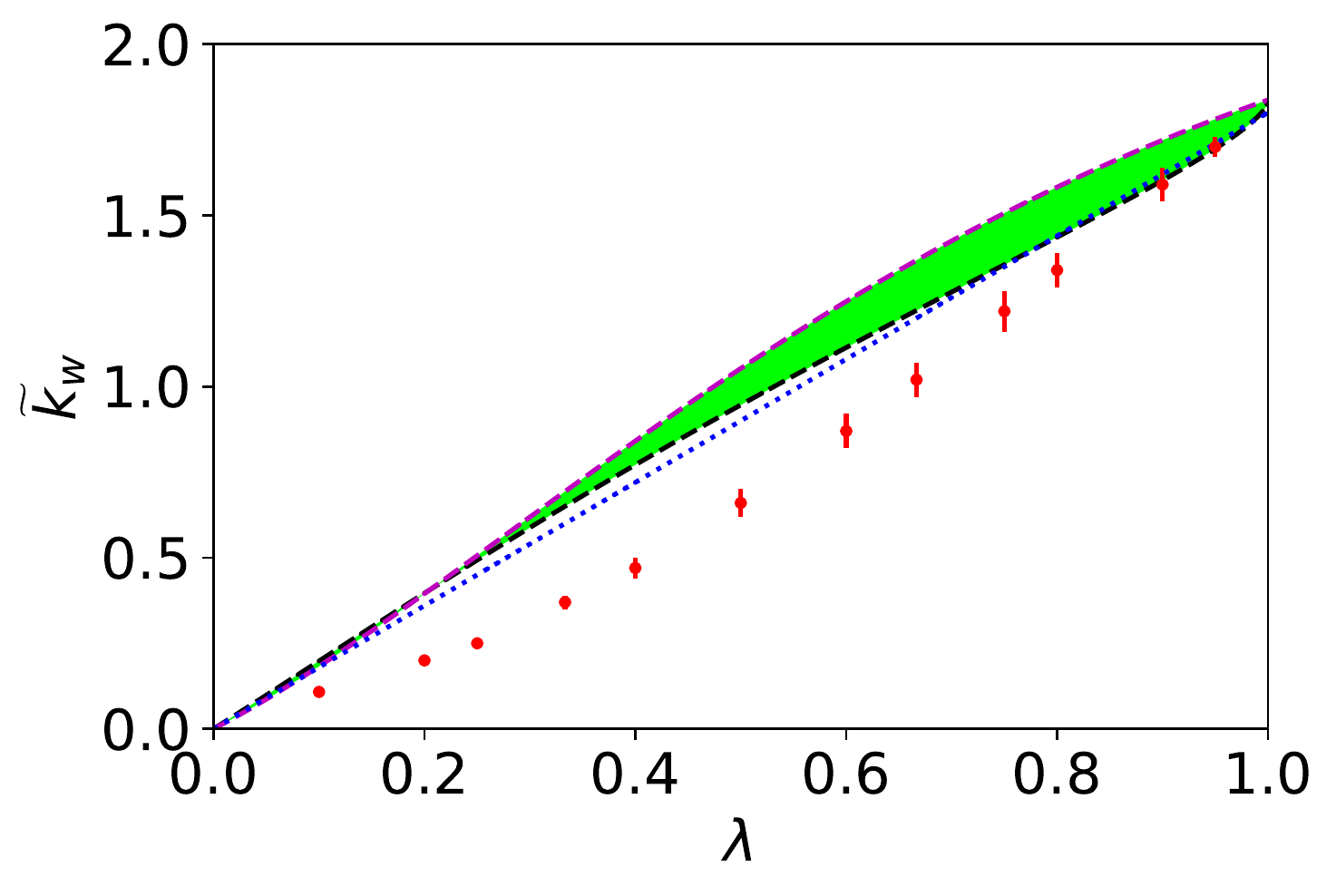}
	\caption{The value of ${\widetilde k}_w$ predicted by our parameter-free VOS model as a function of $\lambda$ in the context of frictionless scaling solutions considering cylindrical or spherical domain walls (upper black and lower magenta dashed lines, respectively) --- the green region between the two lines provides an estimate of the model uncertainty. The red dots with the corresponding error bars represent results obtained using field theory numerical simulations of domain wall network evolution \cite{Martins:2016ois}. Notice that, despite the quantitative disagreement, the parameter-free VOS model and the numerical field theory simulations predict similar qualitative dependencies of ${\widetilde k}_w$ on $\lambda$, which are roughly accounted for by the linear function ${\widetilde k}_w=1.8 \lambda$ (blue dotted line).}
	\label{fig3}
\end{figure}

The frictionless scaling solutions of eqs. (\ref{vos1}) and (\ref{vos2}) for a power law evolution of the scale factor ($a \propto t^\lambda$, $0<\lambda <1$) satisfy
\bq
{\widetilde k}_w  &\equiv& \frac{k_w-\epsilon \zeta}{f+1} = 3 {\mathcal H} \eta  
\zeta \sigma_v = \frac{3\lambda \zeta \sigma_v}{1-\lambda}\,, \label{kwovos}\\
c_w &=&   \frac{\zeta}{   \sigma_v} \left(1  -3 {\mathcal H} \eta  \sigma_v^2\right)= \frac{\zeta \left(1-\lambda(1+3 \sigma_v^2)\right)}{   \sigma_v(1-\lambda)} \label{cwovos}\,.
\eq
Hence, given a scaling solution with $\zeta={\rm const}$ and $\sigma_v={\rm const}$ obtained either using our parameter-free domain wall VOS model or field theory numerical simulations of domain wall network evolution, it is always possible to obtain the values of ${\widetilde k}_w$ and $c_w$ which match that solution. Nevertheless, in the standard parametric VOS model both $f$ and $\epsilon$ are neglected and, therefore, it does not distinguish between $k_w$ and ${\widetilde k}_w$. However, the fact that
\be
\frac{k_w-{\widetilde k}_w}{k_w}=\frac{1}{1+f}\left(f-\frac{\epsilon \zeta}{k_w}\right)\,,
\ee
in combination with the results shown in Figs. \ref{fig1} and \ref{fig2}, shows that, except in the non-relativistic limit, $k_w$ is always significantly smaller than ${\tilde k}_w$.

\begin{figure}[!htb]
	\centering
	\includegraphics[width=12.0cm]{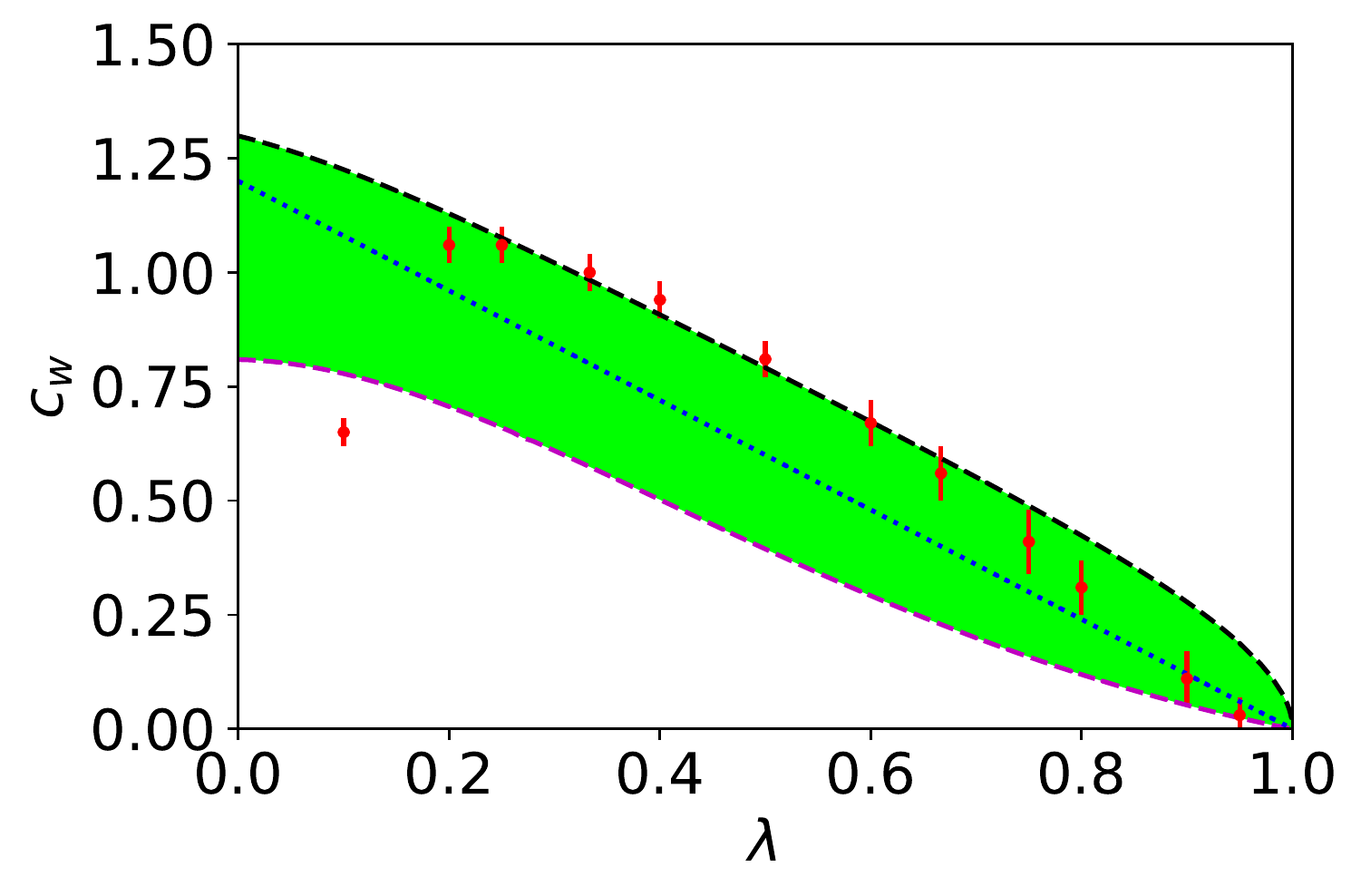}
	\caption{The value of $c_w$ predicted by our parameter-free VOS model as a function of $\lambda$ in the context of frictionless scaling solutions  considering cylindrical or spherical domain walls (upper black and lower magenta dashed lines, respectively) --- the green region between the two lines provides an estimate of the model uncertainty. The red dots with the corresponding error bars represent results obtained using field theory numerical simulations of domain wall network evolution \cite{Martins:2016ois}. Notice that the parameter-free VOS model and the numerical field theory simulations  predict similar qualitative dependencies of $c_w$ on $\lambda$, which are roughly accounted for by the linear function $c_w=1.2(1- \lambda)$ (blue dotted line).}
	\label{fig4}
\end{figure}

Figures \ref{fig3} and \ref{fig4} show, respectively, the values of ${\tilde k}_w$ and $c_w$, as a function of $\lambda$, obtained using eqs. (\ref{kwovos}) and (\ref{cwovos}) with $\zeta$ and $\sigma_v$ given by  our parameter-free VOS model considering cylindrical or spherical domain walls (upper black and lower magenta dashed lines, respectively) --- the green region between the two lines provides an estimate of the model uncertainty. The red dots with the corresponding error bars represent results obtained using field theory numerical simulations of domain wall network evolution \cite{Martins:2016ois}. Here, we have fixed the value of $\beta$ in the cylindrical or spherical domain wall cases by requiring the model to reproduce the value of $\zeta$ obtained using field theory simulations in the non-relativistic limit \cite{Rybak2018} ($\lambda=0.9998$). This requirement gives $\beta_{\rm cylindrical} \sim 0.69$ and $\beta_{\rm spherical} \sim 1.15$ --- note that $\beta$ is independent of the cosmological model and, therefore, it is not a dynamical parameter of our model.

Figure \ref{fig3} shows that, despite the disagreement between the predictions of our parameter-free VOS model and numerical simulations for $\lambda < 0.9$, in both cases the overall dependence of ${\tilde k}_w$ on $\lambda$ may be roughly approximated by the linear function $k_w=1.8 \lambda$, represented by the blue dotted line. Potential causes of the discrepancies observed in Fig. \ref{fig3}, mainly associated to the determination of the $L$ and $\sigma_v$ from field theory numerical simulations, have been discussed in detail in \cite{Avelino:2019wqd} and will need to be tackled in future numerical work.

On the other hand, Figure \ref{fig4} shows that there is overall agreement on the predicted value of $c_w$ between the predictions our parameter-free VOS model and numerical simulations, except for $\lambda=0.1$. The overall dependence of $c_w$ on $\lambda$ may be roughly approximated by the linear function $c_w=1.2 (1-\lambda)$, represented by the blue dotted line.

\section{Conclusions}
\label{sec5}

In this paper we performed a detailed comparison between a recently proposed parameter-free VOS model and the standard parametric VOS model for the cosmological evolution of standard domain wall networks. We have shown that the standard VOS model overestimates the strength of the Hubble damping of wall motion by up to $30 \%$ and neglects the direct impact of wall decay on the evolution of the root-mean-square velocity of the network. We have  also demonstrated how  these effects may be absorbed into a redefinition of its momentum parameter. We have shown that analogous approximations also affect the standard parametric cosmic string VOS model and we have found an additional contribution to the evolution of the root-mean-square velocity of cosmic string networks associated to the impact of loop production. We compared the values of the energy loss and momentum parameters of the standard parametric VOS model for domain walls predicted by our parameter-free one-scale model with those obtained using numerical field theory simulations of domain wall evolution, and provided a simple linear function which approximates their dependence on $\lambda$. 

\acknowledgments

P.P.A. is grateful to Lara Sousa for many enlightening discussions. P.P.A. acknowledges the support from Fundação para a Ciência e a Tecnologia (FCT) through the Sabbatical Grant No. SFRH/BSAB/150322/2019 and through the research grants UID/FIS/04434/2019, UIDB/04434/2020 and UIDP/04434/2020. This work was also supported by FCT through national funds (PTDC/FIS-PAR/31938/2017) and by FEDER—Fundo Europeu de Desenvolvimento Regional through COMPETE2020 - Programa Operacional Competitividade e Internacionaliza{\c c}\~ao (POCI-01-0145-FEDER-031938).


%
%
%
\bibliographystyle{JHEP}
\bibliography{paper}



\end{document}